# Surface Deposition and Imaging of Large Ag Clusters Formed in He Droplets

Evgeny Loginov, Luis F. Gomez, and Andrey F. Vilesov

Department of Chemistry, University of Southern California, Los Angeles, CA 90089

03-18-2011

**Abstract**

The utility of a continuous beam of He droplets for the assembly and surface deposition of $Ag_N$ clusters, $\langle N \rangle \sim 300 - 6\,000$, is studied with transmission electron microscopy. Images of the clusters on amorphous carbon substrates obtained at short deposition times have provided for a measure of the size distribution of the metal clusters. The average sizes of the deposited clusters are in good agreement with an energy balance based estimate of $Ag_N$ cluster growth in He droplets. Measurements of the deposition rate indicate that upon impact with the surface the He-embedded cluster is attached with high probability. The stability of the deposited clusters on the substrate is discussed.

## 1. Introduction

The He droplet technique has proven pivotal in a number of important findings[1-3]. The observation of rotational spectra of molecules embedded in He droplets has provided for a novel microscopic probe of superfluidity in He droplets and its development as a function of droplet size[4-5]. Another very fruitful application of the He droplet technique is in the growth and study of atomic and molecular clusters. Successful coupling of the He droplet technique with laser spectroscopy in the infrared and visible spectral regions in the study of the structure and dynamics of small molecular and atomic clusters in He droplets is reviewed in Refs.[1-3,6-7]. Moreover, He droplets offer straightforward control over the cluster size and allow the formation of multi-component clusters, such as metal-molecule clusters. So far, many cluster experiments with He droplets have employed droplets of less than about $10^4$ atoms resulting in small clusters of no more than about 10 particles. However, previous works indicate that large He droplets can be used to form large atomic or molecular clusters[2]. In particular, the formation of silver and lead clusters of up to about 100 atoms[8-10] and magnesium clusters[11] of up to several thousands of atoms in He droplets has been proven by mass spectroscopic experiments. We showed that ammonia clusters containing up to $10^4$ molecules can be formed in He droplets and studied via infrared spectroscopy[12]. The results of our recent laser spectroscopic study of silver clusters, $Ag_N$ ($N \sim 10 - 10^4$), suggest a transition from single-center to multi-center aggregation in going from small to large He droplets[13]. Thus aggregation in liquid He can also be used to form unique metal samples of nano-granular structure.

Our recent work indicates that metal clusters produced in He droplets can be deposited on a surface upon impact[14]. Supported metal nanoclusters have large potential in heterogeneous catalysis[15], plasmonics[16-17] and single molecule spectroscopy[18]. Of fundamental importance in these areas are the production of size-selected clusters, their controlled deposition onto different substrates, and the stability of such arrays of clusters[19]. In this work, $Ag_N$ clusters formed in He droplets have been deposited on an amorphous carbon (aC) film and studied via transmission electron microscope (TEM) imaging. We will first briefly describe the experimental setup. Secondly, we report on the TEM imaging of $Ag_N$ clusters with estimated average sizes $<N_{Ag}>$ ~ 300 and 6 000 deposited on aC films at various deposition times and on their size distributions and fluxes. Lastly, we discuss the cluster size distributions as well as the mechanism of surface deposition for such clusters grown in He droplets.

## 2. Experiment

The schematic of the molecular beam apparatus is shown in Figure 1. Helium nanodroplets with an average size of $<N_{He}> = 4\cdot10^7$ and $2.4\cdot10^6$ are formed by expanding high purity (99.9999%) He gas at a pressure of 20 bars into vacuum through a 5 μm diameter nozzle at temperatures of $T_0 = 7$ and 9 K, respectively[3]. The beam is collimated by a 0.5 mm diameter skimmer and passes through a 6 cm long differentially-pumped pickup cell at 26 cm from the He droplet source. The pick-up cell contains a resistively heated alumina oven filled with metallic Ag. Further downstream the doped droplet beam enters the deposition chamber where it collides with substrates placed 93 cm from the He droplet source. The substrates are 3 mm diameter standard TEM supports (Ted Pella 01820). They consist of an amorphous carbon film, 15-25 nm thick, mounted on a 300 mesh copper grid coated on the underside by a 30-60 nm thick Formvar film. Typically, a set of 6 samples mounted onto the linear motion manipulator were kept under $10^{-8}$ mbar high vacuum for approximately 24 hours before deposition experiments. The samples were then removed from vacuum and TEM imaging was carried out within 12 hours following deposition. The imaging was conducted on a JEOL JEM-2100 using an electron beam energy of 200 keV. The TEM images were analyzed with the ImageJ image processing package[20].

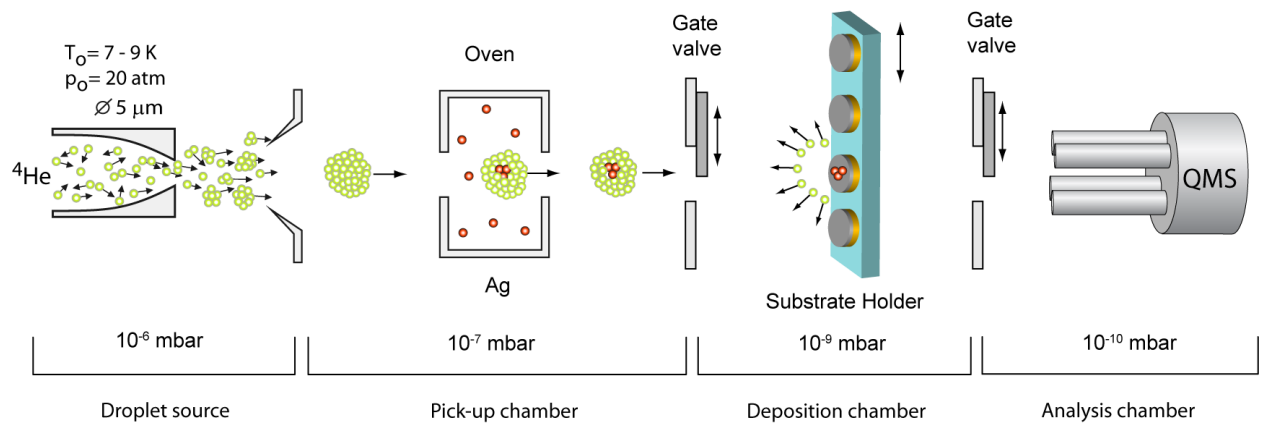

Figure 1. Experimental setup for the surface deposition of metal clusters formed in He droplets. Typical pressure in each vacuum chamber, with the He beam off, is shown.

The average number of Ag atoms captured per He droplet, $<N_{Ag}>$, has been estimated using the attenuation of the He droplet beam, as described in detail elsewhere[14]. The flux of He atoms transported by the droplets is monitored as a rise in the partial pressure of He, $P_{He}$, in the UHV analysis chamber downstream from the deposition chamber which was typically in the range of $10^{-8} – 10^{-7}$ mbar. Upon repeated capture of Ag atoms the average size of the droplets decreases by evaporation of He atoms, which is monitored by a decrease in the pressure rise, $\Delta P_{He}$. The dominant contribution to the energy release upon capture comes from the binding energy of the Ag atoms to the pre-existing $Ag_N$ cluster, and thus $<N_{Ag}>$ can be obtained as:

$$<N_{Ag}> = \frac{\Delta P_{He} \cdot <N_{He}>}{P_{He}} \cdot \frac{E_{He}}{E_{Ag}} \quad (1)$$

where $E_{He}$ is the 0.6 meV[21] binding energy of He atoms to the droplet, $<N_{He}>$ is the initial average size of the He droplet, and $E_{Ag} \approx 3$ eV is the energy associated with the addition of one Ag atom[22].

Maximum flux of silver transported by the He droplet beam is achieved through a balance between doping as many Ag atoms into the droplets while retaining as many carrier droplets as possible. It is known that smaller droplets are extinguished at smaller pickup pressures than larger ones. For a given pickup pressure of silver, a range of smaller He droplets in the size distribution will be fully evaporated and the released $Ag_N$ clusters will be scattered away from the beam axis. Scattering of the free clusters is more efficient, due to their smaller mass, as compared with the clusters embedded in He droplets. Thus, most of the studied clusters must be carried by the He droplets. At the typical doping conditions there are essentially no bare He droplets in the beam. In this work, $Ag_N$ clusters were obtained using an approximately constant attenuation of the He droplet beam of $\Delta P_{He}/P_{He} = 0.7$, which gives the maximum flux of embedded atoms[14]. At the two average droplet sizes employed, $<N_{He}> = 2.4 \cdot 10^6$ and $4 \cdot 10^7$, eq 1 gives $<N_{Ag}> \approx 300$ and 6 000, respectively.

## 3. Results

Figure 2 shows TEM images of samples exposed to the He droplet beam produced at $T_0 =$ 7 K doped with Ag clusters of $<N_{Ag}> \sim 6000$ atoms, as estimated from eq 1, for 0.5 min, 2 min, 32 min and 120 min in panels (a), (b), (c) and (d), respectively. It is seen that the coverage of the samples increases with exposure and at the highest coverage in panels (c) and (d) some clusters are elongated in shape. Such clusters likely result from aggregation of deposited clusters. Figure 2e shows the results of a control experiment in which the sample was exposed for approximately 32 min to an effusive beam of Ag atoms from the oven, which was set to the same temperature as in Figures 2a–d, i.e. with the He droplet beam off. Sample (e) reveals a high density of small clusters of $<N_{Ag}> \approx 200$ formed by aggregation of Ag atoms on the surface. We have observed that such small clusters are not stable under the illumination of the TEM electron beam as they disappear after about 5 min of imaging at a current density of 100 nA/cm$^2$. Apparently, the electron beam imparts the clusters with sufficient temperature such that they either drift out of the field of view or evaporate completely. Previous high resolution studies have indicated a change in the structure of the clusters upon electron beam irradiation[23-24]. This sets a limit on the small cluster sizes which can be imaged in this work at $<N_{Ag}> \approx 200$. It can be estimated that the contribution of the effusive beam is negligible at short exposure times such as 0.5 min, as in panel (a). At longer exposure time, single atoms originating from the effusive beam must combine with large clusters deposited via He droplets or combine into small clusters such as in panel (e).

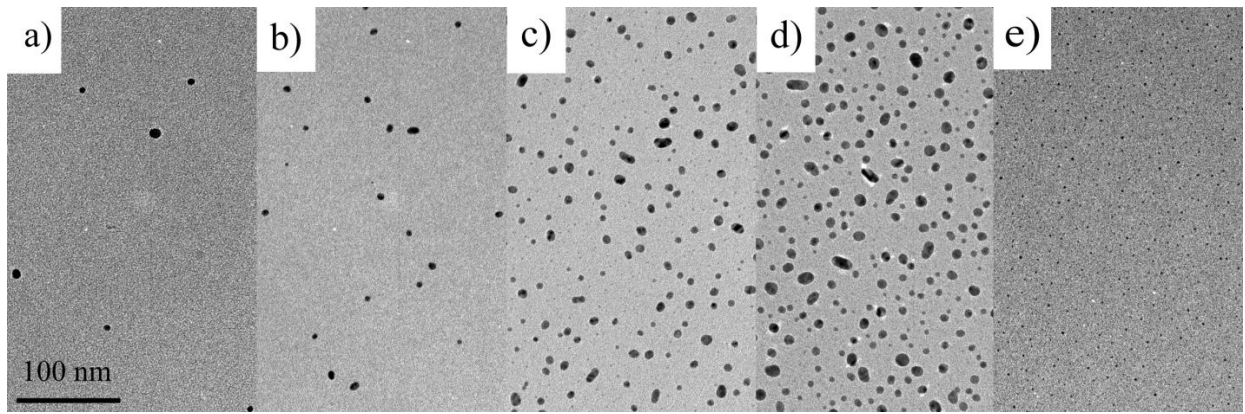

Figure 2. TEM images (at $40 \cdot 10^3$ magnification) of $Ag_N$ clusters on an amorphous carbon film. Samples were obtained by exposure to the He droplet beam doped with about 6000 Ag atoms for 0.5 min (a), 2 min (b), 32 min (c), and 120 min (d). For comparison, panel (e) shows clusters formed on the carbon surface upon 32 min exposure to the effusive beam of Ag atoms emanating from the oven, kept at the same temperature as in experiments producing samples in panels (a) – (d). Scale for all images is the same as shown in panel (a).

For each of the four deposition times shown in Figure 2, a large number of images were recorded to obtain the surface coverage and density of the deposited clusters. The obtained surface coverage and density for each of these is shown in the lower and upper panels of Figure 3, respectively. It is apparent from the insets in Figure 3 that at short deposition time the number of clusters as well as the surface coverage rise almost linearly with time; these, however, become less than linear at longer deposition time, which indicates coalescence of the clusters. The field of view of the electron microscope is of the order of 1 $\mu m^2$, which is much smaller than the diameter of the He droplet beam of about 6 mm at the substrate location . Therefore we do not expect any inhomogeneity of the cluster flux or size distribution over the micrographs. Our results indicate that clusters deposited over a short exposure time of less than about 2 min remain intact and reflect the primary distribution of the deposited clusters.

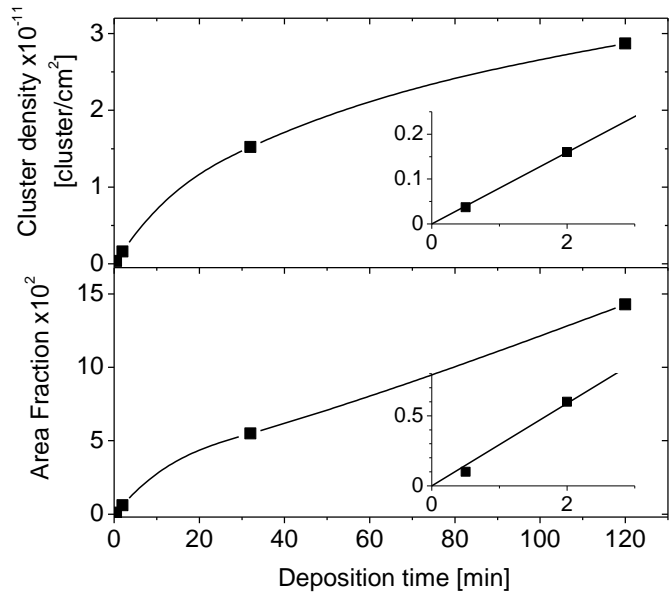

Figure 3. Upper panel: Surface density versus deposition time for Ag$_N$ clusters of estimated initial size $<N_{Ag}>$ ~ 6 000. Line connecting the data points is to guide the eye. Lower panel: Area fraction of the deposited clusters under the same conditions. Insets show a linear fit to the data at short deposition times.

For an imaged cluster of surface area $S$, we can calculate the number of Ag atoms, $N_{Ag}$, contained within as:

$$N_{Ag} = \frac{4}{3}\pi^{-\frac{1}{2}}S^{\frac{3}{2}}n, \qquad (2)$$

where $n = 5.85 \cdot 10^{22}$ cm$^{-3}$ is the number density of bulk silver. Here, we assume a spherical shape of the deposited clusters. Small clusters, as in this work, must be nearly spherical because the Ag-Ag$_N$ binding energy of $\approx 3$ eV[22] is stronger than that of the Ag-carbon surface of $\approx 1$ eV[25].

Based on the results in Figure 3, new samples with deposition times of 2 min and 32 min were prepared for analysis of the cluster size distribution. The samples were imaged at a magnification of $40 \cdot 10^3$. Approximately 1 000 clusters were detected over 50 images taken of different areas of the 2 min sample. Each image was obtained within less than a minute to minimize any possible distortion of the size distribution induced by the TEM electron beam. The obtained size distributions are shown in Figure 4 by filled squares and open circles for the 2 min and 32 min samples, respectively. The mean cluster size obtained according to eq 2 from the 2 min samples was $<N_{Ag}> = 6\ 400$ with a root mean square deviation $\Delta N_{Ag} = 5\ 000$. In comparison, the distribution of the clusters obtained at the longer deposition time of 32 min, where clusters with $N_{Ag} > 15\ 000$ are abundant, is shifted towards larger sizes. As a result, the mean cluster size at 32 min deposition is almost a factor of two larger with $<N_{Ag}> = 11\ 800$ and $\Delta N_{Ag} = 11\ 400$. This must be a result of aggregation of the clusters at high surface density, in agreement with the results in Figure 3.

From the surface density of the clusters at short deposition time, such as in Figure 2b, the silver deposition flux was obtained to be $7 \cdot 10^{11}$ clusters/(sr·s) or $8 \cdot 10^7$ clusters/(cm$^2$·s). The corresponding atomic flux is $4.4 \cdot 10^{15}$ atoms/(sr·s) or $5 \cdot 10^{11}$ atoms/(cm$^2$·s). The cluster flux is a factor of two smaller than the flux of undoped He droplets of $1.6 \cdot 10^{12}$ droplets/(sr·s) estimated from the pressure rise in the UHV analysis chamber and average He droplet size. The rate of cluster formation from the effusive beam with a mean cluster size of $<N_{Ag}> \approx 200$ atoms, such as in Figure 2e, was obtained to be $4.7 \cdot 10^7$ clusters/(cm$^2 \cdot$s). This corresponds to an effusive Ag flux of $9 \cdot 10^9$ atoms/(cm$^2 \cdot$s), which is a factor of 100 smaller compared to that transported in He droplets. This comparison shows that Ag$_N$ clusters in He droplets are the primary source of the deposited clusters in Figure 2a–d, whereas the effusive beam is relatively weak.

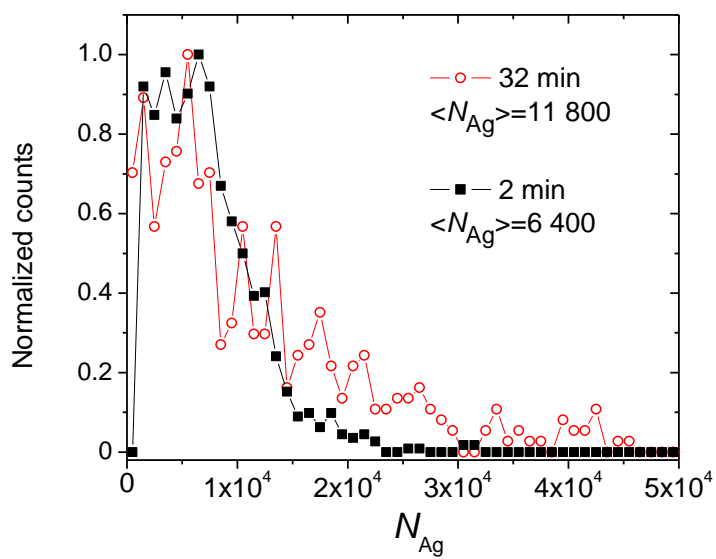

Figure 4. Size distribution of Ag$_N$ deposited on an aC film during 2 minutes (solid squares, 1 000 clusters analyzed) and 32 minutes (open circles, 400 clusters analyzed). Estimated initial average cluster size in both cases is 6 000. Lines connecting data points are to guide the eye.

Additional experiments were carried out at an increased He droplet source temperature of 9 K to deposit smaller Ag$_N$ clusters, estimated at about 300 atoms from eq 1, via He droplets consisting of about 2.4·10$^6$ atoms. Figure 5 shows the size distributions obtained after 2 minutes and 16 minutes of deposition, indicated by stars and open circles, respectively. The mean size of the deposited clusters produced in He droplets was obtained to be <$N_{Ag}$> = 600, $\Delta N_{Ag}$ = 600 for deposition over 2 min and <$N_{Ag}$> = 800, $\Delta N_{Ag}$ = 700 for deposition over 16 min. Thus the obtained sizes are about a factor of two larger than estimated. On the other hand, the distributions of the deposited clusters peak at $N_{Ag} \approx 400$ in better agreement with the estimate. The average cluster size at longer deposition time is only somewhat larger indicating that the deposition flux for 16 min is still low enough to avoid substantial cluster aggregation (area fraction is less than 1%). In addition, the appearance of the rather long tail at larger sizes (up to $N_{Ag}$ = 6 000, not shown in Figure 5) in the distribution for 16 min may indicate that smaller clusters remain mobile upon deposition and some of them coalesce even at low coverage. From the cluster counts in Figure 5, the deposition flux was obtained to be 5·10$^{12}$ clusters/(sr·s) or 5.6·10$^8$ clusters/(cm$^2$·s). The corresponding atomic flux is 3·10$^{15}$ atoms/(sr·s) or 3.4·10$^{11}$ atoms/(cm$^2$·s). The Ag$_N$ cluster flux is about a factor of three smaller than the estimated flux of undoped He droplets of 1.7·10$^{13}$ droplets/(sr·s). The obtained value, however, is in very good agreement with the previously measured deposition flux of Ag atoms, under similar experimental conditions[14] employing a quartz crystal microbalance, of 3.2·10$^{15}$ atoms/(sr·s).

The distribution of clusters, shown in Figure 2e, resulting from the effusive beam (without He droplets) over an exposure of 32 min is shown in Figure 5 by solid squares. It is seen that clusters obtained from the effusive beam are smaller, with a distribution peaking at $N_{Ag} \approx 200$. The distribution may be somewhat biased towards larger sizes as we are not able to reliably detect clusters of less than $N_{Ag} \approx 200$ or of diameter smaller than about 2 nm due to low contrast in the TEM images, as well as cluster evaporation under the electron beam illumination.

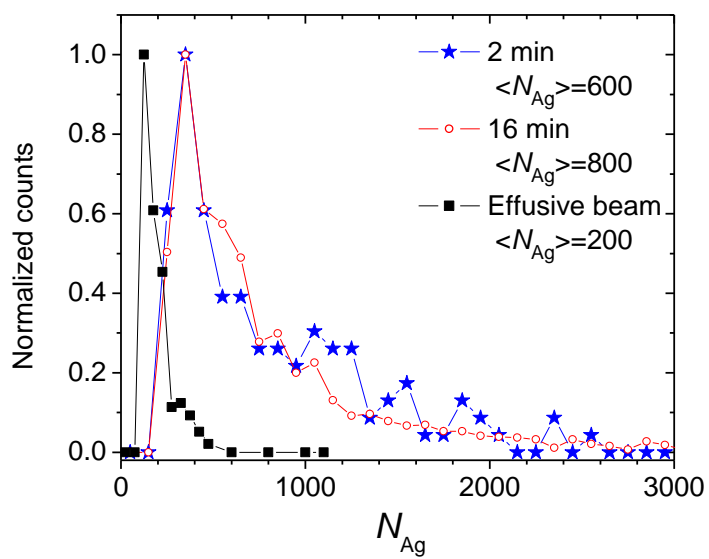

Figure 5. Size distribution of Ag$_N$ clusters deposited on an aC film for 2 minutes (stars, 200 clusters analyzed) and 16 minutes (circles, 2 000 clusters analyzed). Estimated initial average cluster size in both cases is 300. Squares illustrate the cluster size distribution upon exposure to the effusive beam for 32 minutes at the same conditions as in Figure 2e. Lines connecting data points are to guide the eye.

## 4. Discussion

The ratio of the flux of deposited $Ag_N$ clusters to that of He droplets in the beam gives the deposition yield of the clusters. Our results show that the flux of $Ag_N$ clusters determined from the TEM images is about a factor of two to three smaller than the flux of He droplets. This may indicate a somewhat smaller than unity sticking probability for the clusters in He droplets. On the other hand, a difference of this magnitude may be due to uncertainties in the estimates. For example, the initial He droplet sizes in this work may differ from that in previous measurements[3,26] due to the use of a different nozzle plate and some slight inaccuracy in the actual nozzle temperature. In addition, according to Ref.[27], there is a considerable and unknown fraction of small droplets in the beam which adds uncertainty to the estimates based on average droplet size. Moreover, the measurements of the droplet size distribution via deflection of droplets having an attached electron in Ref.[28], which we rely on, cannot be used to study droplets of less than about $10^5$, since such droplets do not bind an electron. Therefore, we conclude that the sticking probability of $Ag_N$ clusters in He droplets colliding with the aC surface is large but cannot be determined with high accuracy at present. In comparison, in our previous work[14], where the Ag flux was measured by a microbalance, both cluster flux and the droplet flux were equal within experimental error: $7.5 \cdot 10^{12}$ clusters/(sr·s) and $7 \cdot 10^{12}$ droplet/(sr·s), respectively.

### 4.1. Helium droplet collision with surface

The cold He droplets eventually disintegrate following collision with the warm surface, releasing the embedded $Ag_N$ clusters. The surface collisions of microscopic droplets (e.g., water, ethanol, etc.) has been thoroughly investigated via fast photography as well as theoretical calculations; see for example reviews in Refs. [29-30] and references therein. The experimental conditions used in this work such as nanometer-sized He droplets, nano-scale substrate roughness, and high vacuum conditions differ substantially from those in the reviews. However, for estimation purposes, we consider the results obtained for such macroscopic droplets. It is well known that the outcome of a collision such as rebounding, spreading, or splashing depends on the collision speed $v$, density within the droplet $\rho$, droplet diameter $D$, and surface tension of the liquid $\sigma$, as well as on properties of the substrate. The initial phase of a collision of an incompressible drop depends on the magnitude of the Weber number, $We$, given by:

$$We = \frac{\rho v^2 D}{\sigma} \qquad (3)$$

It is seen that $We$ is proportional to the ratio of kinetic energy to the surface energy. Bouncing of drops is observed at small Weber number, $We < 10$, and depends on wetting of the surface by the drop liquid. Usually, splashing occurs when $We$ exceeds a certain critical value of about $We = 100$; see for example Ref.[31].

In this work, He droplets have an impact velocity of about 200 m/s, the helium density at 0.4 K is 145 kg/m$^3$,[32] and the surface tension is $3.54 \cdot 10^{-4}$ N/m [32]; thus the magnitude of $We$ is estimated to be 1600. In addition, collisions with surfaces at temperatures above the critical temperature of the liquid are influenced by heat transfer and the fast evaporation of the droplet liquid. In the present experiments, the substrate temperature of about 300 K is substantially higher than the critical temperature of liquid helium, 5.2 K. Experiments with classical droplets (water and organic liquids) having values of $We \approx 2\,000 - 10\,000$ were reported by Pan $et\ al.$[33]. Manzello and Yang[34] have studied collisions at surface temperatures above the critical point of the drop liquid and $We = 700 - 750$. In both studies, fast imaging of the surface impact of mm-sized water droplets shows considerable spread of the droplet during the impact leading to the formation of a liquid disc of diameter up to about ten times larger than the initial droplet. If the same scenario is valid for He droplets of 100 nm diameter, as employed in this work, containing Ag$_N$ clusters as in Figure 2, the resulting thickness of the liquid disc will be around 1 nm which is smaller than the diameter of the embedded $Ag_{6000}$ cluster of about 6 nm. Of course, fracture of the disc may occur at some point when its height becomes comparable to the distance between He atoms in the liquid of about 0.4 nm as follows from the number density of liquid He of 21.8 $nm^{-3}$.[32] Thus we expect that during the impact the dopant $Ag_N$ cluster will come in direct contact with the aC surface and remain attached to it, while the He droplet disintegrates. This scenario is in agreement with the observed high yield of the deposited clusters. During the expansion along the surface of the colliding droplet its rim can attain a high velocity comparable to the velocity of sound in liquid He of about 240 m/s[32]. Therefore, the embedded clusters may also be dragged along the surface before attachment. This He-assisted mobility of the clusters on the surface may contribute to enhanced combination of the clusters. At present, it remains unclear whether heat transfer from the surface will be important during the short time of the expansion phase following collision which lasts for about 0.5 ns.

**4.2. Soft landing**

The $Ag_N$ clusters in He droplets collide with the aC at the velocity of the droplet beam which is known to be about 200 and 300 m/s for the larger and smaller droplets[35] used in this work, respectively. Within the assumption that the kinetic energy of the clusters is determined by the velocity of the carrier droplets, the kinetic energy per impacting Ag atom in the large clusters is about 0.034 eV. This is much less than one tenth the binding energy of a single Ag atom to the cluster of about 3 eV and to the surface which is about 1 eV for amorphous carbon[25]. Thus He droplet deposition is well within the so called "soft landing" regime[19]; accordingly, the cluster and the surface remain intact upon collision. The estimated collision energy is less than the lowest yet reported in the literature of 0.05 eV for $Sb_N$ ions, $<N> = 90 - 2\ 200$, impacting $aC^{36}$.

### 4.3. Cluster size distribution

The size distribution of the Ag$_N$ clusters is a convolution of both the pick-up probabilities and the He droplet size distribution in the beam. In the supercritical expansion regime, as used in this work, the droplet size decays approximately exponentially towards larger sizes, having a mean square deviation comparable to the mean size: $\Delta N_{He}/\langle N_{He}\rangle \approx 1$.[26] Thus the width of the distribution of the neat He droplet sizes is comparable to that for deposited Ag clusters, where we found $\Delta N_{Ag}/\langle N_{Ag}\rangle \approx 0.8$. Therefore we conclude that the width of the cluster size distribution is mainly defined by the size distribution of the hosting He droplets. There must also be some bias in our experiment towards larger droplets which carry large clusters. Smaller Ag$_N$ clusters are formed in smaller He droplets from the distribution; such droplets are not only more effectively evaporated in collisions with Ag atoms but also scattered more. Both of these effects combine as a bias towards larger clusters.

The final size distribution, shape, and morphology of the deposited clusters are defined by the interplay between the dynamics of the doped He droplet impact[29-30], the nature of the substrate surface, cluster aggregation[37-39], secondary processes[40], and the TEM imaging procedure itself[23-24]. As discussed earlier, Ag$_N$ clusters may experience some considerable translation along the surface during the droplet impact prior to adsorption, contributing to the mobility of the clusters on the surface and increasing the possibility of cluster aggregation. This depends on, for instance, the deposition rate[37-39] and the kinetic energy of the impacting clusters[41]. Once deposited, adsorption is facilitated by the known high density of surface defects of aC that serve as adsorption sites for deposited metal clusters[37]. The long term mobility of the clusters depends on the nature of these adsorption sites. Clusters will either remain pinned to some strong adsorption site or diffuse over the substrate until they find a suitable adsorption site or combine with another Ag$_N$ cluster. In fact, clusters on an atomically flat surface are known to remain mobile and can agglomerate by diffusion to form larger metal islands which are fractals under certain conditions[39]. The linear dependence of the surface coverage at short deposition time in the present experiments shows that aggregation of clusters is not important at surface densities of less than about $1.6 \cdot 10^{10}$ clusters/cm$^2$. However, this becomes important at higher surface densities. The largest surface density of clusters in Figure 3 corresponds well to the saturation cluster density of around $5 \cdot 10^{11}$ clusters/cm$^2$ measured for various metal clusters (In, Bi, Au, Ag) on aC films[37-38,40,42].

The obtained mean cluster size of $<N_{Ag}>$ = 6 400 in Figure 4 is in good agreement with the estimate based on the binding energy of atoms in Ag$_N$ clusters and the initial average size of the He droplets. Thus we conclude that clusters remain intact upon deposition and that aggregation of the clusters is not important at low coverage, such as in Figure 2a-b. Despite this conclusion, clusters may experience some reconstruction upon attachment to the surface at room temperature. Recently, we have studied the structure of Ag$_N$ clusters formed in He droplets (*in situ*) via optical spectroscopy[13]. We have found that in smaller clusters, such as the $N_{Ag}$ ~ 600 clusters studied in this work, the spectra are dominated by a surface plasmon resonance near 3.8 eV consistent with absorption by individual compact metallic particles. However, larger clusters, such as the $N_{Ag}$ ~ 6400 clusters studied in this work, reveal unexpectedly strong broad absorption at low frequency extending down to ≈0.5 eV. This suggests a transition from single-center to multi-center formation, in agreement with estimates of cluster growth kinetics in large He droplets. Accordingly, a number of small clusters , which later coalesce into an aggregate inside the hosting He droplet, are formed during pickup. These small clusters may retain their individuality because the energy barrier associated with reconstruction into a close packed cluster is insurmountable at 0.4 K in He droplets. However, upon deposition onto a surface at room temperature cluster aggregates must coalesce into compact particles. The coalescence must be facilitated by the long residency of the clusters on the surface of half a day prior to TEM imaging, including several hours at ambient conditions, as well as by exposure to the electron beam during the imaging procedure. This conjecture is supported by the results of Ref.[39], which show that small ($N < 1\,000$) Ag$_N$ clusters deposited on an aC surface aggregate and coalesce within a few hours.

The average cluster size of smaller clusters, $<N_{Ag}> = 600$ in Figure 5, is a factor of two larger than that estimated from eq 1, which may indicate some aggregation of the clusters. It may also be the result of low contrast in the TEM images of small clusters, such that the obtained distribution is biased towards larger sizes. In fact, it has been previously observed that Au clusters of less than 1 nm diameter could not be detected immediately after deposition due to low contrast in the TEM images[38,43]. Popescu et al.[40] reported the detection of Au clusters of 0.3 nm diameter only after 7 days exposure to ambient atmospheric conditions, which may indicate the importance of some secondary processes resulting in increased contrast in the TEM images of small clusters. The electron beam itself, requisite for TEM imaging, is also known to induce evaporation of small deposited Ag$_N$ clusters. This would also explain the inability to reliably image clusters of less than about 200 atoms in the present work. Finally, the accuracy of the estimates according to eq 1 heavily rely on the average He droplet sizes from Refs.[3,26]. The He droplet sizes obtained at the nominal nozzle temperature of 9 K in this work strongly depend on the *actual* nozzle temperature; any slight variation in the actual nozzle temperature would lead to droplets of initial average size different from those reported in Refs.[3,26]. Therefore, the Ag$_N$ clusters assembled within will likewise exhibit a strong size-dependence on the actual nozzle temperature adding to any uncertainty in the estimates. Therefore, the disagreement factor of about two between the estimated and measured cluster sizes is not unexpected.

## 5. Conclusions

We have demonstrated that metal clusters formed in He droplets can be successfully extracted by deposition onto an amorphous carbon film. The mean size of the deposited $Ag_N$ clusters is within about a factor of two in agreement with the estimated initial cluster size, which is based on the binding energy of the clusters and initial He droplet size. This shows that such clusters remain intact upon deposition and TEM imaging. This study opens up new opportunities for routinely synthesizing in He droplets diverse metal and metal-molecule clusters spanning a wide range of cluster sizes and compositions and for studying their deposition on different substrates.

## Acknowledgements

We are grateful to undergraduate student Caroline Lin for her help with some of the experiments and data analysis described herein. This material is based upon work supported by the NSF under grant CHE 0809093. This paper is dedicated to Prof. J. P. Toennies on the occasion of his 80$^{th}$ birthday.